\documentclass[twoside]{dis09}
\usepackage[latin1]{inputenc}
\usepackage{graphicx,color}
\usepackage{wrapfig,rotating}
\usepackage{amssymb,amsmath,array}

\begin{document}
\title{Strategies to Extract Generalized Parton Distributions from Data}

%***********************************************************************
% AUTHORS INFORMATION AREA
%***********************************************************************
\author{Simonetta Liuti, Saeed Ahmad, Chuanzhe Lin and Huong T. Nguyen
%
% Optional short acknowledgment: remove next line if non-needed
\thanks{U.S. Department
of Energy grant no. DE-FG02-01ER4120}
%
% DO NOT MODIFY THE FOLLOWING '\vspace' ARGUMENT
\vspace{.3cm}\\
%
% Addresses and institutions (remove "1- " in case of a single institution)
University of Virginia - Physics Department \\
382 McCormick Rd., Charlottesville, VA 22904 USA }
%
% Remove the next three lines in case of a single institution
%\vspace{.1cm}\\
%2- Tufts University - Department of Physics and Astronomy \\
%Medford, MA 02155 - USA\\
%%***********************************************************************
% END OF AUTHORS INFORMATION AREA
%***********************************************************************

\maketitle

\begin{abstract}
A number of deeply virtual exclusive experiments will allow us to access the Generalized Parton Distributions which are embedded in the complex 
amplitudes for such processes. The extraction from experiment is particularly challenging both because 
of the large number of kinematical 
variables and observables  to be pinned down in each experimental analysis and because, 
at variance with inclusive experiments, the variables
representing the quark momentum fraction appear integrated over in the physical amplitudes and cannot be accessed directly.
We present a strategy for the extraction from experiment that makes use of constraints from both elastic and inclusive scattering as well as 
information from lattice QCD results.   
\end{abstract}

%***************************************************************************
\section{Introduction}
Deeply virtual exclusive experiments such as $e p \rightarrow e^\prime \gamma p^\prime$ 
(Deeply Virtual Compton Scattering, DVCS) and $e p \rightarrow e^\prime M p^\prime$ (Deeply Virtual Meson Production, DVMP) can provide information on the partons' localization in space in addition 
to their longitudinal momentum fraction distribution \cite{RalPir}.
The scattering amplitude for DVCS can be written within QCD factorization at leading order in $1/Q$
as
\begin{equation}
\label{DVCS}
T^{\mu\nu} =-\frac{1}{2}g^{\mu\nu}_T \, {\bar u}(p^\prime){\hat n}u(p)
\sum_{q} e_q^2 {\cal F}_q (\xi,t),
\end{equation}
where $Q^2$ is the four-momentum transfer squared in the hard collision, $p$($p^\prime$) is the initial (final) proton momentum, $t=\Delta^2$ is the four-momentum transfer squared between the initial and final protons, $\xi= 2 \Delta^+/(p+p^\prime)$ is the skewness variable, and   
\begin{equation}
\label{direct}
{\cal F}_q ^+(\xi,t)=\int\limits_{-1}^{+1}dx \frac{F_q ^+(x,\xi,t)}{x-\xi+i\epsilon}.
\end{equation}
The GPD $F_q(x,\xi,t)$ (with $F_q=H_q$ or $E_q$) is the soft part in the handbag diagram describing this reaction, which is 
convoluted with the hard part,  $1/(x-\xi+i\epsilon)$, and integrated over $x$, representing the 
partons longitudinal momentum fraction. 
Crossing symmetry is implemented by
\begin{equation}
F_q^{(\pm)}(x,\xi,t)=F_q(x,\xi,t) \mp F_q(-x,\xi,t),
\end{equation}
recalling that for PDFs, $q(-x)=-{\bar q}(x)$ relates negative $x$ to positive $x$ antiquark probability.

The goal of this contribution is to determine a physically motivated GPD parametrization that 
satisfies known theoretical constraints from both elastic and inclusive scattering as well as 
information from lattice QCD results. The formulation that we present ultimately aims at: ``devising 
a form combining essential dynamical elements with a flexible model that allows for a fully 
quantitative analysis constrained by the data".  
%A first study performed in \cite{AHLT1,AHLT2}    
It is a step forward in the direction of current global analyses in related sectors such as the 
extraction of Transverse Momentum Distributions (TMDs) from semi-inclusive experimental data
\cite{Anselmino,BacRad}.
Differently from TMDs however the extraction of GPDs is expected to be more challenging  due to
the large number of kinematical 
variables and observables  to be pinned down in each experimental analysis. In addition, as illustrated in Eq.(\ref{direct}),
at variance with inclusive and semi-inclusive experiments, the variables
representing the quark momentum fraction appear integrated over in the physical amplitudes and cannot be accessed directly.
Other features also appear that are unique to GPDs, and should be taken care of in a physically motivated parametrization, such as  $Q^2$ evolution (see {\it e.g.} discussion in \cite{DM_proc}),   
the interplay between flavor separation and crossing symmetry \cite{DM_proc,GG_proc,GL_disp},
the ability to constrain a given GPD from the limited number of available measurements 
\cite{GuiMou}, and finally the applicability of dispersion relations \cite{GG_proc,GL_disp}.
   
Here we discuss results 
obtained using the phenomenologically constrained 
parametrization from Refs.\cite{AHLT1,AHLT2}, and we present further 
developments of the model. The parametrization is valid for unpolarized GPDs 
in the valence sector, at intermediate values
of the skewness. It is aimed at having the same kinematical coverage of Jefferson Lab 6 GeV and 12 GeV experimental programs. 

Our formulation is based on a quark-diquark picture improved by a Regge-type contribution (for more details see \cite{AHLT1,AHLT2}:

\begin{equation} 
\label{param1} 
 H(X,\zeta,t) = G(X,\zeta,t) R(X,\zeta,t),
\end{equation}
where $G$ and $R$ are respectively the quark-diquark, and Regge-type contributions, and we dropped the subscript $q$ for simplicity. Note that the parametrization is given in terms of the longitudinal variables $X=k^+/p^+$, and $\zeta=\Delta^+/p^+$, which are in a one to one correspondence with the alternative set $x,\xi$ (see \cite{Diehl_rev} for a review).  The parametrization is constructed at an initial
scale $Q_o^2$, whose value is determined by our  fit to the inclusive and elastic scattering data. 
$H$ can subsequently be evolved perturbatively \cite{Rad_pert}.  

Experimental constraints can be found by using the fact that $H$ obeys the following relations
\cite{AHLT1}:
\begin{eqnarray}
\label{forward}
H(X,0,0) &  =  & q(X) \\
\label{ff}
\int\limits_{-1+\zeta}^1 \frac{dX}{1-\zeta/2} \, H(X,\zeta,t) &  = & F_1(t),
\end{eqnarray}
$q(X)$ being the (valence) quark distribution, and $F_1$ the Dirac form factor. 
Clearly, these constraints do not affect the $\zeta$ dependence, however additional conditions
can be found by using ab initio lattice QCD calculations of the higher moments 
in $X$ (\cite{Hagler_07} and references therein).  
The $n=1,2,3$ moments for both the isovector and isoscalr combinations: $H_{u-d} = H_u-H_d$, and $H_{u+d} = H_u+H_d$ 
are in fact available (moments for the GPD $E$ are also available but we will not use them in this contribution). Constraints from lattice QCD were used for the first time in Ref.\cite{AHLT2}.

By imposing the above constraints it was established in Ref.\cite{AHLT1} that the quark-diquark picture cannot reproduce data/current parametrizations on $q(X)$ at low $X$.
One can in fact expect  that the quark-diquark or spectator models where the mass of  the spectator is kept at a fixed value 
does not have the right physical input to reproduce the power dependence in $X$ of the data. 
The  term $R(X,\zeta,t)$ takes care of this problem effectively. This in turn allows for a quantitative description of  both the low $X$ behavior, and of the $t$ dependence in the whole range of $t$. It is important to stress  that a correct treatment of the low $X$ behavior is crucial in determining the $t$ dependence of GPDs, even if considering the valence region, because it dominates the form factor sum rule, Eq.(\ref{ff}).  The latter cannot be satisfied without introducing a  Regge type behavior at low $X$. This feature is missing from diquark models. A more thorough study is under way where Regge behavior emerges from a spectral distribution in the proton's soft debris invariant mass, $M_X$, valid at large values of the center of mass energy squared, $s$ \cite{GolLiu_prep}.  

For the present form of our parametrization  we treated separately the so-called DGLAP ($X>\zeta$) and ERBL ($X<\zeta$) regions. 

The functional forms for Eq.(\ref{param1}) in the DGLAP region are given by
\begin{subequations}
\label{param2}
\begin{eqnarray}
\label{diq_zeta}
G(X,\zeta,t)  &  =  &
{\cal N} \frac{X}{1-X} \int d^2{\bf k}_\perp \frac{\phi(k^2,\lambda)}{D(X,{\bf k}_\perp)}
\frac{\phi({k^{\prime \, 2},\lambda)}}{D(X,\zeta,{ \bf k}_\perp^\prime)}, \\ 
\label{regge}
R(X,\zeta,t) & = & X^{- \alpha - \beta (1-X)^p (t+t_{min})},
\end{eqnarray}
\end{subequations}
where 
\[ k^2 =  X M^2 - \frac{X}{1-X} M_X^{2}  - \frac{{\bf k}_\perp^2}{1-X} \]

\[k^{\prime \, 2}  =  \frac{X-\zeta}{1-\zeta} M^2 - 
\frac{X-\zeta}{1-X} M_X^{2} - \left({\bf k}_\perp - \frac{1-X}{1-\zeta} \Delta \right)^2 \frac{1-\zeta}{1-X},
\]
In Eq.(\ref{diq_zeta})$k$ and $k^\prime$ are the initial and final quark momenta respectively, 
$D(X,{\bf k}_\perp) \equiv  k^2 - m^2$, $D(X,\zeta,{\bf k}_\perp^\prime) \equiv  k^{\prime \, 2} 
- m^2$, ${\bf k}_\perp^\prime ={\bf k}_\perp - (1-X)/(1-\zeta)\Delta$, $m$ being the struck quark mass. 
$\phi(k^2,\lambda)$ is the nucleon-quark-diquark vertex function, where the  
diquark is either a scalar or an axial-vector (this allows us to separate out the $d$ and $u$ quarks components).
In Eq.(\ref{regge}) 
$t_min= -\zeta^2/(1-\zeta)M^2$. 

Notice that the hybrid parametrization does not make use of a
``profile function'' for the parton distributions, 
but the forward limit, $H(X,0,0) \equiv q(X)$, 
is enforced non trivially. This affords us the flexibility that 
is necessary to model the behavior at $\zeta, \, t \neq 0$.
The parameters in the DGLAP region were fitted separately: the set $\{ \alpha, \lambda, M_X \}$ was first used to determine the $\zeta=0, t=0$ behavior, {\it i.e.} fitted to $q(x)$ in the valence region;
next, the set $\{p, \beta\}$ was used to fit to the nucleon form factor data. All parameters are listed in  \cite{AHLT1}. We reiterate that $Q_o^2$, the initial scale,  is also determined by the fit in our approach. More importantly, although all parameters
are determined  in our approach at $\zeta=0$, the $\zeta$ dependence in the DGLAP region is obtained through the kinematics expressed in Eqs.(\ref{param2}). The values of the parameters obtained for $R$ do not correspond directly to similar ones from Regge theory because of the multiplicative form in Eq.(\ref{param1}) that is such that {\it e.g.} the slope at low $X$ receives contributions from both the $R$ and $G$ functions. 
  
To extend a parametrization in the  ERBL region one needs  
$\zeta$-dependent constraints given {\it e.g.} by the higher moments of GPDs that are calculable in {\it ab initio} calculations in
lattice QCD. Our study is dedicated to defining and using these constraints. 

%%%%%
%%%%% Figure
\begin{wrapfigure}{r}{0.5\columnwidth}
\centerline{\includegraphics[width=0.45\columnwidth]{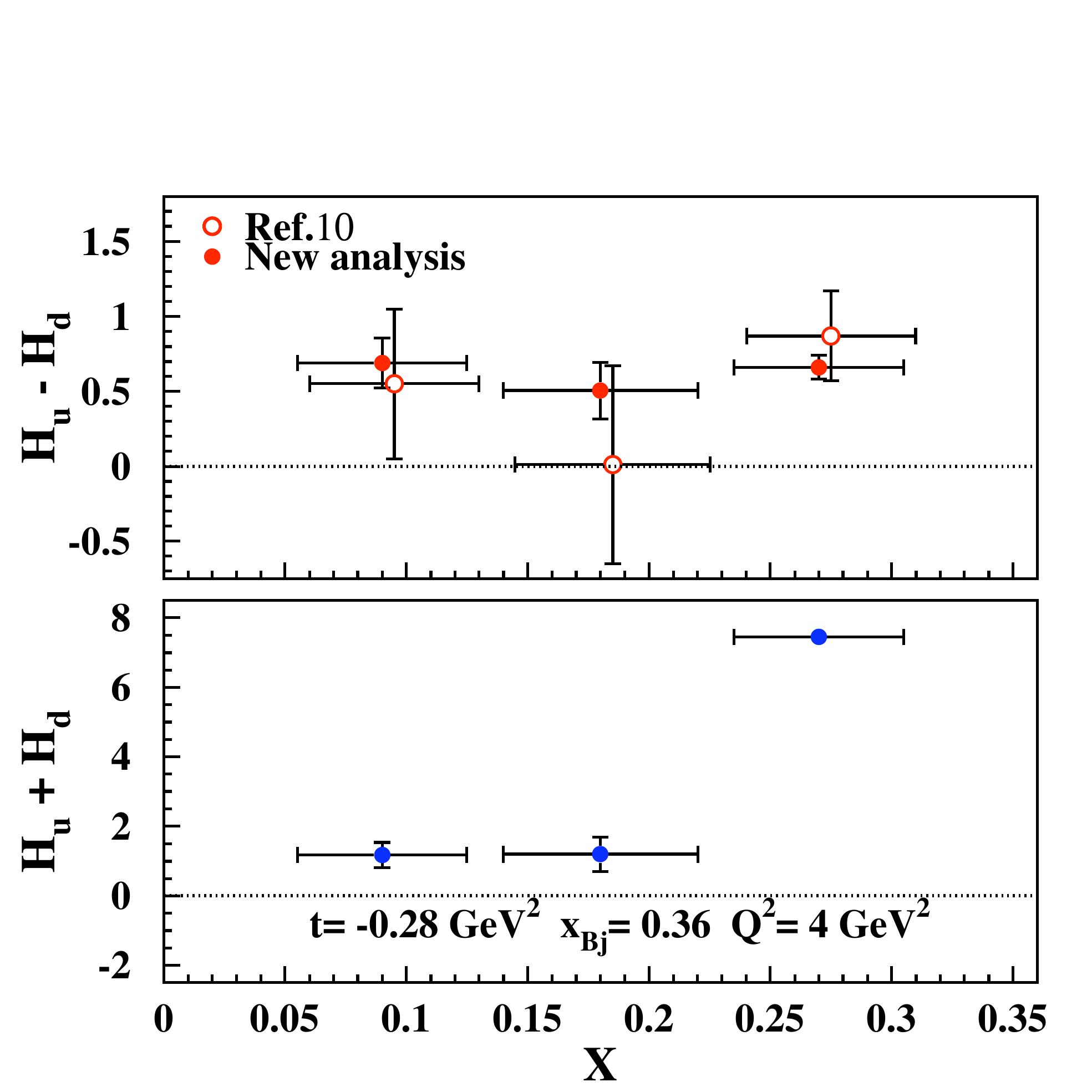}}
\caption{The isovector, $H_u-H_d$ (top),  and isoscalar, $H_u+H_d$ (bottom)  contributions to the generalized parton distribution, $H$ plotted as a function of $X$ at fixed $x_{Bj}= \zeta$, $t$, and $Q^2$, in the ERBL region ($X<\zeta=0.36$) extracted using the lattice QCD calculations from Ref.\cite{Hagler_07}. The errors on the variable $X$ represent the dispersion from the Bernstein moments method \cite{AHLT2}, while the errors on $H$ reflect the error on the lattice results.}
\label{fig1}
\end{wrapfigure}
%%%%%%
%%%%%%

In Fig.\ref{fig1} we present one sample of our preliminary results from a study performed using the most recently available lattice evaluations from Ref.\cite{Hagler_07}. The figure shows $n=3$ points for both the isovector, $H_u -- H_d$ , and isoscalar, $H_u+H_d$, contributions plotted as a function of $X$ at fixed kinematics. The points where obtained by using a mathematical technique first used in 
\cite{Yndurain} by which the function $H(X,\zeta,t)$ is {\em reconstructed} in average ($H \equiv \overline{H}$) over a given set of points, $\overline{X}$ in the 
interval $X \in [0,\zeta]$,
using a set of normalized polynomials, in this case the Bernstein polynomials, as weight functions around $\overline{X}$ ($\zeta$ and $t$ are kept fixed). 
The Bernstein moments, or the weighted averages of $H$, are then obtained as linear combinations of 
$A_1(\zeta,t),  A_2(\zeta,t), A_3\zeta,t)$ which are obtained from the lattice calculations in \cite{Hagler_07}.   
Notice that the latter are moments in the ERBL region, obtained by subtracting the DGLAP region from the total moments given in \cite{Hagler_07}.  
Furthermore, the lattice calculations on GPD moments need to be chirally extrapolated.
In \cite{AHLT2}  we used a phenomenological approach 
that can be applied to all $n=1,2,3$ moments. The approach makes use of the fact that all moments seem to show a dipole behavior displaying increasing values of the dipole masses, $\Lambda^{(n)}$ with increasing $n$.  Chiral extrapolations for the dipole masses are then implemented. In \cite{AHLT2} we checked that results were consistent with both data and alternative methods in the case $n=1$. Here we were able to check that our new results are consistent with the newly available  chiral extrapolations for $n=2$ from Ref.\cite{Dorati}. Our consistency check is important for testing the reliability of our method for
 $n=3$ for which current calculations in chiral perturbation theory are impracticable, while well tested 
 phenomenological approaches like the ones proposed can be used. 
The errors on the $x$-axis represent the dispersion in our reconstruction technique while the errors on the $y$-axis 
 are propagated from the lattice calculation errors. No other source of errors is taken into account. From Fig.\ref{fig1} one can see that the isovector contribution is consistent with previous determinations. One can notice a much improved error bar with respect to our previous analysis in \cite{AHLT2} due to the 
 improved set of lattice results. Furthermore, it  is now possible to calculate also the isoscalar contribution.   
 
The dispersion along the  the $x$-axis could improve if higher $n$ moments could be calculated using different lattice QCD techniques that avoid operator mixings and the associated  renormalization issues \cite{Det}. 
Nevertheless, with better constraints in hand,  such as the ones displayed in Fig.\ref{fig1} showing both reduced errors on the isovector part, and for the first time the isoscalar contribution, one can now consider a fully fledged parametrization in the ERBL region.

In conclusion, we presented first results on a physically motivated parametrization  that takes into 
account a number of constraints on the GPD H in the valence quarks sector. Similar results were also 
obtained for the GPD $E$.
Parameters along with our statistical analysis are made available. 
Our work represents a step forward in the much needed direction of Global Parametrizations of both TMDs and GPDs. 
More comparisons with Jefferson Lab data on both DVCS and DVMP including the chiral-odd GPD sector \cite{AGL} are on their way.

\section*{Acknowledgments}
We thank Gary Goldstein for insightful comments. We are also grateful to Swadhin Taneja and Ross Young
for discussions. This work was funded by the U.S. Department
of Energy grant no. DE-FG02-01ER4120.
 
% ****************************************************************************
% BIBLIOGRAPHY AREA
% ****************************************************************************

\begin{footnotesize}
% IF YOU DO NOT USE BIBTEX, USE THE FOLLOWING SAMPLE SCHEME FOR THE REFERENCES
% ----------------------------------------------------------------------------

% ----------------------------------------------------------------------------

% IF YOU USE BIBTEX,
% - DELETE THE TEXT BETWEEN THE TWO ABOVE DASHED LINES
% - UNCOMMENT THE NEXT TWO LINES AND REPLACE 'Name_Of_Your_BibFile'

%\bibliographystyle{unsrt}
%\bibliography{Name_Of_Your_BibFile}
% example of Name_Of_Your_BibFile.bib
% @Article{Turcato:2006ch,
%      author    = "Turcato, M.",
%  collaboration = "ZEUS and H1",
%      title     = "Lepton flavour violation and charmonium physics at HERA",
%      journal   = "Nucl. Phys. Proc. Suppl.",
%      volume    = "162",
%      year      = "2006", 
%      pages     = "283-287",
%      SLACcitation  = "%%CITATION = NUPHZ,162,283;%%"
% }
% 
% @Unpublished{Gogitidze:2007du,
%      author    = "Gogitidze, N.",
%  collaboration = "H1", 
%      title     = "Prompt photons and particle momentum distributions at
%                   HERA", 
%      year      = "2007",
%      note    = "hep-ex/0701033",
%      SLACcitation  = "%%CITATION = HEP-EX 0701033;%%"
% }

\end{footnotesize}

% ****************************************************************************
% END OF BIBLIOGRAPHY AREA
% ****************************************************************************

\end{document}